\documentclass[review,number,sort&compress]{elsarticle}

\usepackage{amssymb}
\usepackage{amsmath}
\usepackage{units}
\usepackage{url}
\usepackage{booktabs}
\usepackage{multirow}
\usepackage{natbib}
\usepackage{stfloats} 

\usepackage{placeins} 

\journal{Nuclear Physics A}

\begin{document}

\begin{frontmatter}




 \title{Coincident Detection of Cherenkov Light from Higher Energetic Electrons Using Silicon Photomultipliers}
 \author[address-label]{Reimund Bayerlein} 
 \author[address-label]{Ivor Fleck}
 \author[address-label]{Waleed Khalid}
 \author[address-label-2]{Todd E. Peterson}
 \author[address-label]{Albert H. Walenta}

 \address[address-label]{Center for Particle Physics Siegen, University of Siegen, Walter-Flex-Stra{\ss}e 3, 57072 Siegen, Germany\fnref{label3}}
 \address[address-label-2]{Vanderbilt University Institute of Imaging Science and the Department of Radiology and Radiological Sciences, Vanderbilt University Medical Center, Nashville, TN 37212, United States}

\begin{abstract}

Due to their very fast signal rise time in the order of 1\,ns, Silicon-Photo\-multi\-pliers have become of increasing interest for many experiments that require very good timing resolution. With the prospect of an application in medical imaging techniques like a Compton Camera or TOF-PET, a coincident detection of Cherenkov photons from electrons in the MeV range in PMMA has been performed. A $4\times 4$ SiPM-array was used for this purpose and a timing resolution of 242\,ps has been achieved. A spatial sensitivity for an electron source location could be shown using accumulated coincident light signals. Obtained results are in good agreement with theoretical calculations taking fundamental detector and set-up properties into account. These measurements constitute an important step towards the feasibility of a successful electron detection in a Compton Camera.

\end{abstract}


\end{frontmatter}


\section{Introduction}
\label{sec-introduction}

In recent years silicon photomultipliers (SiPM) have become commonly available and are employed in a multitude of applications. 
Especially in medical applications SiPMs are used because of their very fast signal rise time and consequently good timing resolution. 
One example would be time-of-flight positron-emission-tomography (TOF-PET) \cite{ref:Vandenberghe}. 
A further improvement of the timing resolution and consequently the position resolution can be achieved when using Cherenkov photons in addition to the scintillation light~\cite{ref:brunner}.

There are, however, other applications in medical imaging where Cheren\-kov radiation could play a role: The detection of Cherenkov photons from Compton scattered electrons is one possible concept that could help realize a Compton Camera \cite{ref:peterson}.
The aim of a Compton Camera is the detection and the reconstruction of the momentum direction of high energetic photons ($\geq$\,1\,MeV) produced e.g. in targeted alpha therapy or particle beam therapy. 
In this energy range Compton scattering is the dominant process for the photon interaction. 
The detection concept comprises a two layer system, where in the first one the incident photon scatters 
releasing an energetic Compton electron. In a second layer the scattered photon is then absorbed. 
The Compton electron carries a large fraction of the momentum information of the incident photon. Thus, position and energy sensitive detection of the electron in coincidence with the scattered photon allows the reconstruction of the incident direction of the photon to lie on the surface of a cone. This can be further reduced to the segment of a cone if the electron momentum direction is measured as well \cite{ref:Roellinghoff}. 

One possible concept for the electron detection is the coincident measurement of Cherenkov photons caused by that electron. 
These photons are generated along the electron track and radiated under a characteristic angle, forming the so called Cherenkov cone. 
Assuming an electron speed close to $c$, the opening angle is mainly determined by the refractive index of the material (See equation \ref{eq:cos-beta}). 
Reconstruction of that cone yields information on the interaction vertex of the Compton scattering and also on the momentum direction of the electron, while the number of detected photons contains information on the electron energy \cite{ref:peterson}.
Multiple scattering of the electron degrades the cone structure, however, simulations suggest that spatial information can still be extracted \cite{ref:ota, ref:peterson2016}.

For an electron with a kinetic energy of 1\,MeV Cherenkov radiation is emitted within a time window of less than 20 ps for a non-scintillating radiator material. 
To be able to achieve sufficient timing resolution for the coincidence detection on the sub-nanosecond scale, SiPMs are a promising choice for a photon detector. 

As a proof of principle, the coincident detection of Cherenkov photons from electrons in PMMA (poly-methyl-metacrylate) using a $4\times 4$-SiPM array will be demonstrated in this paper. The influence of the thickness of the PMMA sample on the width of the distribution of Cherenkov photons as well as on the number of created photons is investigated. A coincidence time resolution on the order of 250\,ps is achieved and a reconstruction of the electron source position from accumulated coincident events is demonstrated.



The paper is organized as follows: First the expected Cherenkov photon yield for the used experimental setup is calculated in section~\ref{sec:theory} while the setup itself is described in section~\ref{sec:setup}. 
Next the performed measurements are given in section~\ref{sec:measurements} and the results are presented in section~\ref{sec:results} and are discussed with respect to the theoretical expectations.
\FloatBarrier
\section{Theoretical Background and Calculations}
\label{sec:theory}

\subsection{The Cherenkov Effect}
\label{subsec:chkv-effect}
Whenever a charged particle exceeds the speed of light $c_\mathrm{n}$ in that medium, electromagnetic radiation is emitted \citep{ref:Cherenkov34}. Asymmetric polarization of the dielectric medium and subsequently the coherent superposition of elementary waves creates so-called Cherenkov radiation, which is emitted along the surface of a cone pointing towards the direction of flight. One key parameter of this effect is the characteristic opening angle $\theta$ of this cone with respect to the direction of electron movement, depending only on the speed $v$ of the particle and the refractive index $n$:
\begin{equation}\label{eq:cos-beta}
\cos \theta = \frac{c_\mathrm{n}}{v} = \frac{c}{n\cdot v} = \frac{1}{\beta\,n} = \frac{1}{n}\cdot \left[ 1-\left( \frac{m_0c^2}{m_0c^2+E_\mathrm{kin}} \right)^2 \right]^{-\frac{1}{2}},
\end{equation}
where $m_0$ is the particle's rest mass and $E_\mathrm{kin}$ is its kinetic energy.
Here, $\beta = v/c$ denotes the velocity of the particle as a fraction of the vacuum speed of light $c$. 
Above a threshold of $\beta_\mathrm{th} = 1/n$, Cherenkov light is produced 
and the opening angle increases with $\beta$ up to a maximum of $\cos \theta = 1/\mathrm{n}$.
For Poly-methyl-metacrylate (PMMA) with a refractive index of 1.49 this threshold is at $\beta_\mathrm{th} = 0.67$, which in case of electrons is equivalent to a kinetic energy of 177\,keV.
Unlike scintillation light, the emission of Cherenkov light happens instantaneously all along the electron track.
\begin{figure}
\begin{center}
\hspace*{-0.3cm}
\includegraphics[width=0.8\textwidth]{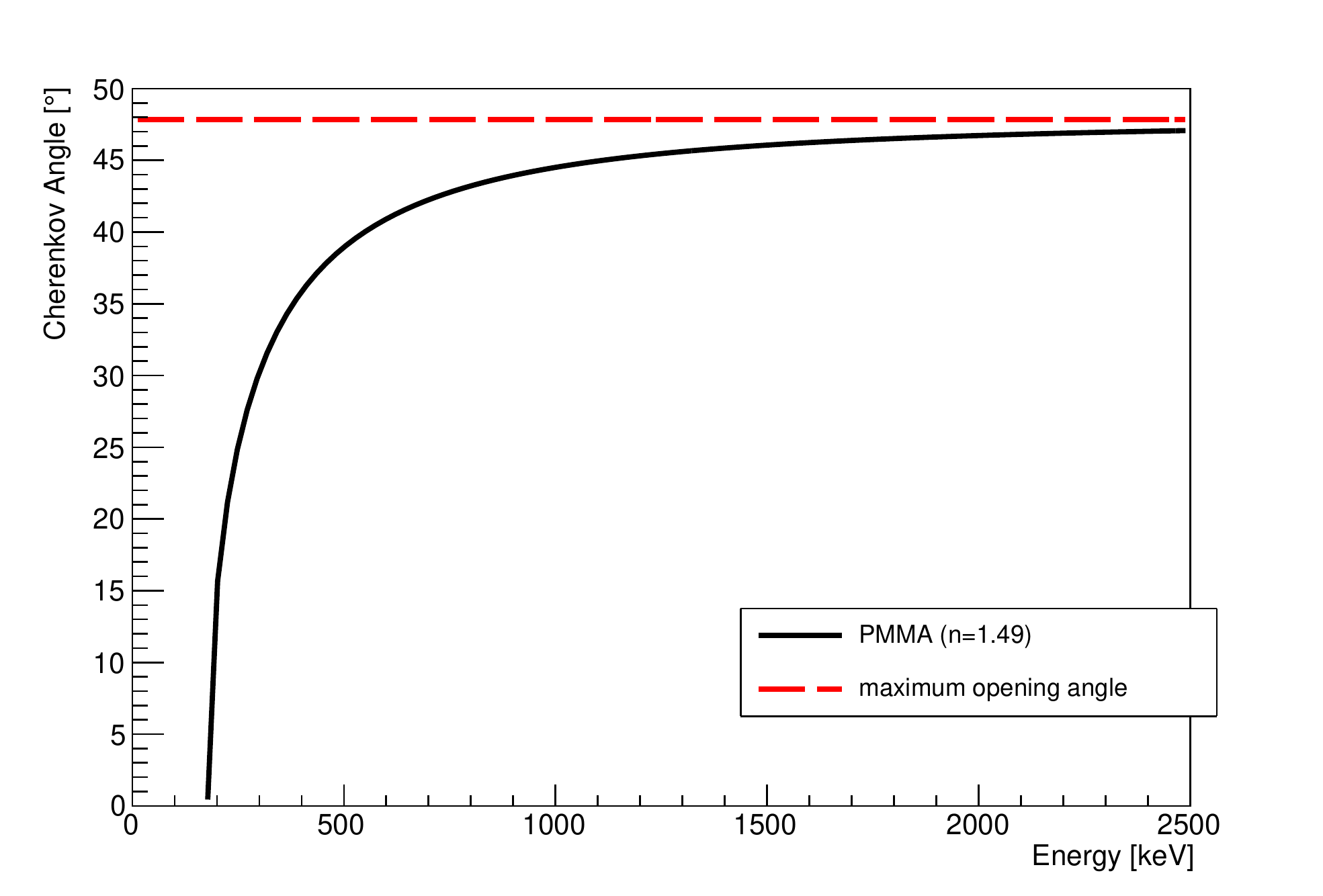}
\caption{Cherenkov cone angle versus kinetic energy of the electron in PMMA ($n = 1.49$) based on equation \ref{eq:cos-beta}.}
\label{fig-cone-angle}
\end{center}
\end{figure}

Figure \ref{fig-cone-angle} shows the opening angle of the Cherenkov cone in PMMA 
as a function of the kinetic energy of the electron according to formula \ref{eq:cos-beta}. 
The angle grows rapidly with increasing energy and converges towards its maximum. 
At 780\,keV the angle has already reached 90\,\% of its maximum value, 
which for PMMA is 47.8\,${}^\circ$. 
Due to its energy dependence, the opening angle of the Cherenkov cone gets smaller throughout the electron track in the medium as the energy of the electron decreases.
In addition, the number of emitted photons becomes more sparse with decreasing energy 
(compare section \ref{subsec:num-gen-ph} and especially equation \ref{eq:dN^2/dldx}). 

\subsection{Number of Generated Cherenkov Photons}
\label{subsec:num-gen-ph}

\begin{figure}
\begin{center}
\includegraphics[width=0.8\textwidth]{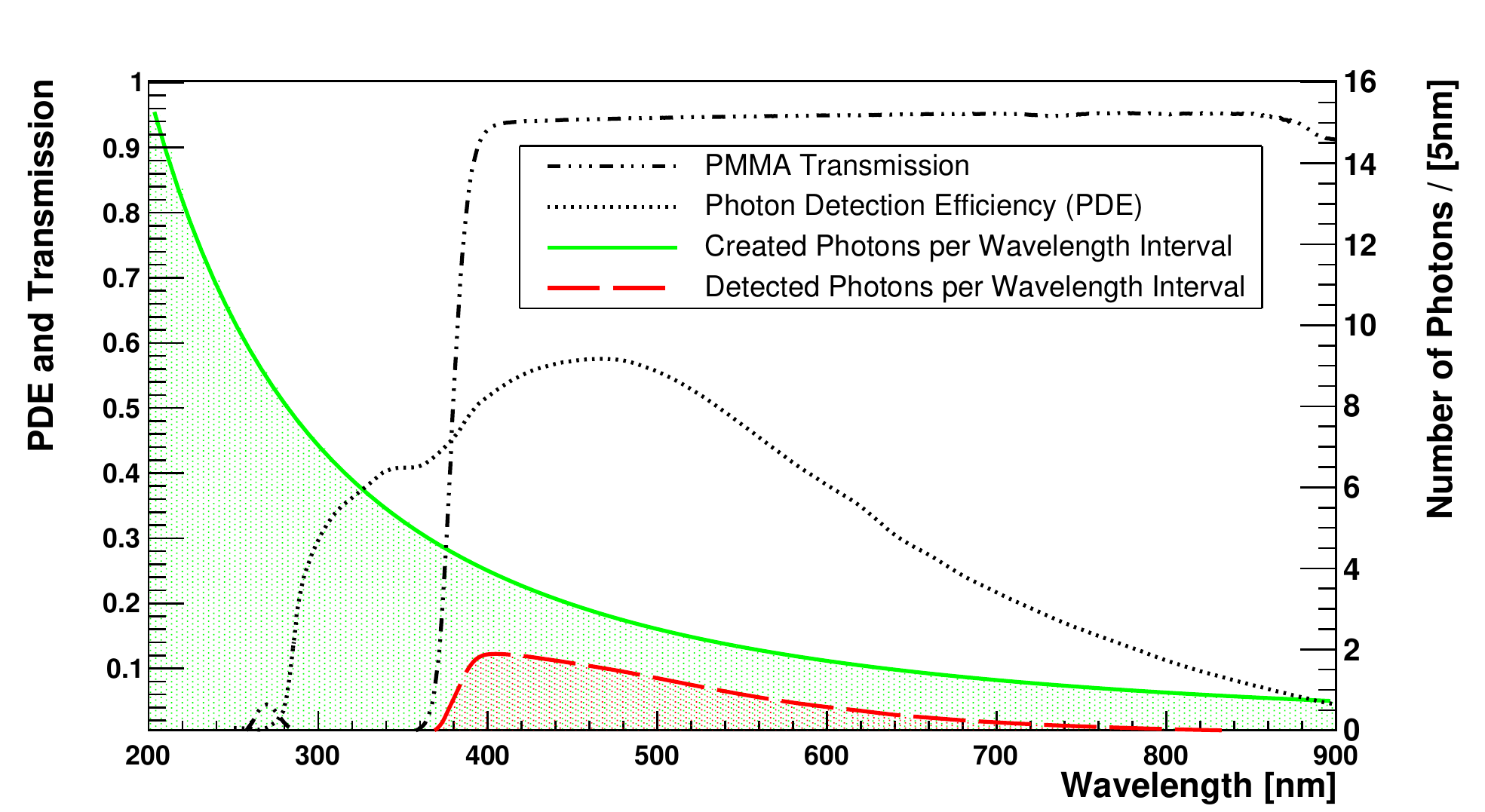}
\caption{Theoretical estimations on the number of detected Cherenkov photons (long dashed line) from a 1.5\,MeV electron taking into account light transmission of PMMA \cite{ref:sultanova} and bias-dependent detection efficiency of the SiPM \cite{ref:hamamatsu}. The solide curve shows the number of created photons per wavelength interval of 5\,nm based on equation \ref{eq:N_i}.}
\label{fig-multiplot}
\end{center}
\end{figure}

The number of generated Cherenkov photons within a wavelength interval d$\lambda$ emitted along some travelled distance d$x$ can be written as \cite{ref:kolanoski}
\begin{equation}\label{eq:dN^2/dldx}
\frac{\mathrm{d}N^2}{\mathrm{d}x\mathrm{d}\lambda} = \frac{2\pi \, z^2\,\alpha}{\lambda^2}\cdot \left( 1-\frac{1}{\beta^2\,n^2(\lambda)} \right).
\end{equation}
This is also referred to as \textit{differential Cherenkov photon emission}, where $\alpha$ denotes 
the fine structure constant, $\alpha= \unitfrac{1}{137}$, and $z$ is the charge of the particle in units of $e=1.6 \cdot 10^{-19}$\,C, which in the case of electrons equals 1.
The refractive index $n(\lambda)$ is wavelength dependent, but for PMMA changes are on the order of 1\,\% 
in the visible range~\cite{ref:sultanova}. 
As formula~\ref{eq:dN^2/dldx} indicates, the number of emitted photons per traveled distance and per wavelength interval increases with the energy of the electron. The expression becomes zero for $v=c_\mathrm{n}$ and is approximately constant for large energies. The dependency on $1/\lambda^2$ causes an increased photon yield towards smaller wavelengths. Photon detectors, therefore, need to have good efficiency in the near UV range in order to achieve high efficiency for the detection of Cherenkov light. Figure \ref{fig-multiplot} shows that wavelength dependency of the number of created Cherenkov photons (solid line). The graph displays the number of created photons per wavelength interval for a 1.5\,MeV electron in PMMA. 

\begin{figure}
\begin{center}
\includegraphics[width=0.8\textwidth]{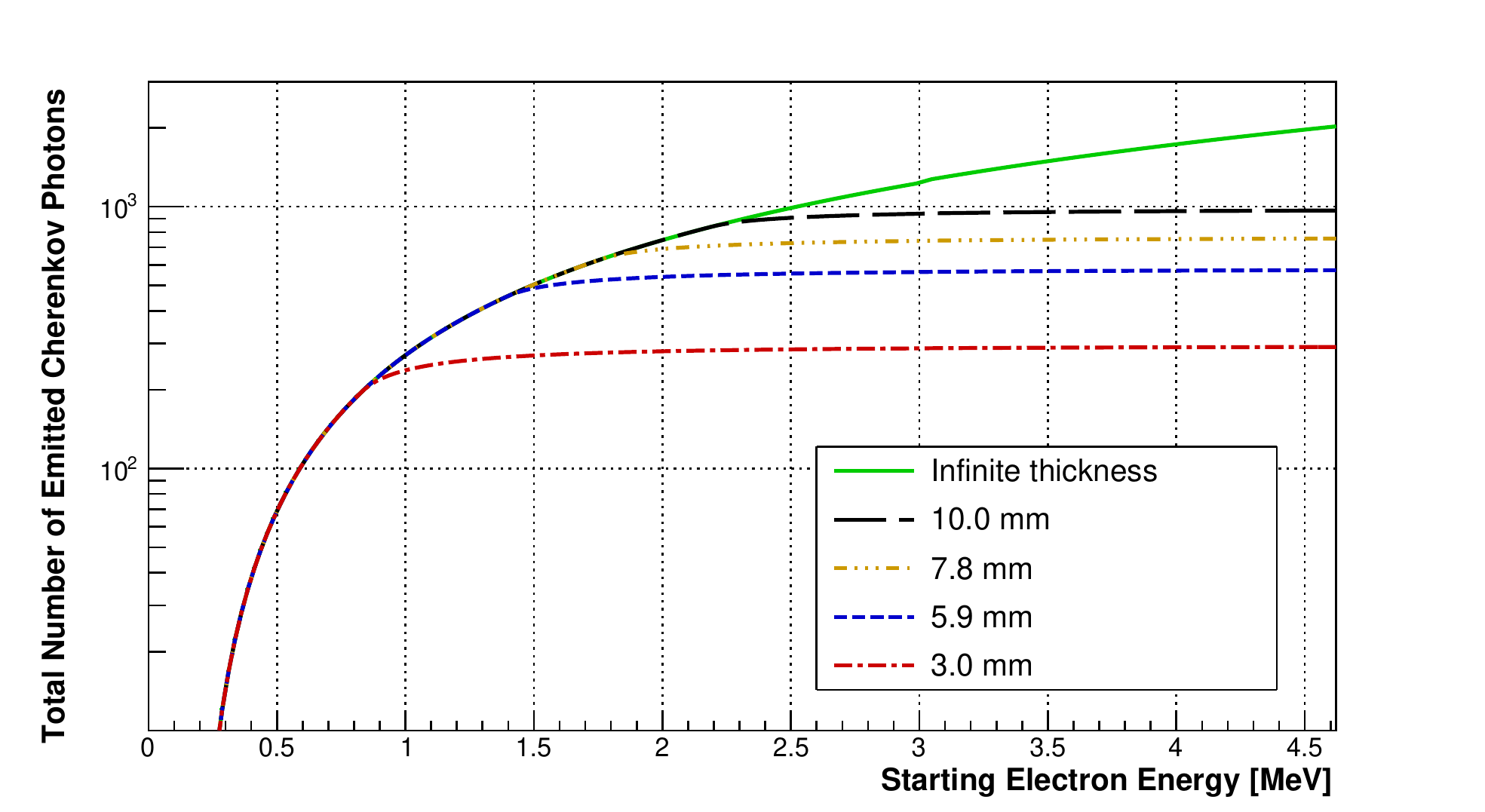}
\caption{Calculated number of generated Cherenkov photons in PMMA in the wavelength range between 200\,nm and 900\,nm depending on the electron energy. Results are shown for different thicknesses of PMMA. Thinner samples limit the range of the electron and therefore the light yield, causing the number of created photons to saturate.}
\label{fig-numPh-diff-thick}
\end{center}
\end{figure}

The underlying calculation uses \textit{Stopping Power and Range Tables for Electrons} (ESTAR) from the\textit{ National Institute for Standards and Technology} (NIST) \cite{ref:estar}. For each energy $E$ the table provides the stopping power $\mathrm{d}E/\mathrm{d}x$ for electrons in PMMA. Choosing a small enough step size in energy, one can assume the stopping power to be approximately constant along that step. This can be used to calculate the traveled distance $\Delta x$ of the particle between two energy values:
\begin{equation}
\Delta x = \frac{\Delta E}{\mathrm{d}E/\mathrm{d}x}
\end{equation}
Then, the number of photons emitted within that step is calculated by integrating formula \ref{eq:dN^2/dldx}.
Assuming the refractive index to be constant within the wavelength range of interest, the number of photons in that interval reads
\begin{equation}\label{eq:N_i}
N_i=2\pi \alpha\cdot (\Delta x)_i\cdot \left(1-\frac{1}{n^2\,\beta^2}\right)\cdot\left(\frac{1}{\lambda_1}-\frac{1}{\lambda_2}\right),
\end{equation}
with $\lambda_2 > \lambda_1$.
This calculation is performed for all provided energies above the Cherenkov threshold and all calculated steps $(\Delta x)_i$. 
The wavelength range was chosen between 200 and 900\,nm.
The total number of Cherenkov photons for an electron with a given starting energy is then simply the sum of all partial photon numbers $N_i$ from all steps $(\Delta x)_i$:
\begin{equation}\label{eq:N_tot}
N_\mathrm{tot} = \sum_i	N_i.
\end{equation}
Figure \ref{fig-numPh-diff-thick} shows the total number of produced Cherenkov photons as a function of initial electron energy. This estimation was performed for different thicknesses of the radiator material PMMA. The summation in equation \ref{eq:N_tot} was conducted until the total sum of all steps $\sum_i (\Delta x)_i$ had reached the thickness of the PMMA sample. 
The number therefore saturates at a certain energy, above which the electron track length would be larger than the thickness of the sample. The solid curve shows the number of emitted photons in an infinitely thick sample.

For all mentioned calculations the assumption is made that the electron track is a straight line and multiple scattering (which deflects the electron) happens mostly at the end of the track, where the electron does not have enough energy to maintain the Cherenkov effect. 

\subsection{Estimating the Number of Detected Photons}
\label{subsec:estimation-prog}
To be able to compare obtained measurement results with theoretical expectations, the number of detected Cherenkov photons from electrons in PMMA using a Silicon Photomultiplier (SiPM) was calculated. 
The algorithm calculates the number of created Cheren\-kov photons in the range of 200 - 900\,nm. 
Based on equation \ref{eq:N_i} the spectral distribution of this number follows a $1/\lambda$ relation enhancing the emission for smaller wavelengths.
In order to estimate the number of \textit{detected} photons per wavelength interval each number of created photons is multiplied with the transmission of PMMA and the photon detection efficiency (PDE) of the detector, the latter of which scales with the applied detector bias voltage. 
This voltage dependency was implemented using the manufacturer's statement about the change of the peak PDE with bias voltage, and the PDE of each wavelength interval was scaled accordingly.
Figure \ref{fig-multiplot} shows the transmission of PMMA and the PDE of the used SiPM for an overvoltage of 3.8\,V.
For this case, the number of created photons for an electron with an energy of 1.5\,MeV is 504 while the estimated number of photons detected by the SiPM is 75.
The solid line and the long dashed line show the number of photons per 5\,nm interval that were created and detected, respectively. 
The energy threshold above which the calculated number of Cherenkov photons in the wavelength range between 200\,nm and 900\,nm is at least 1 equals 201\,keV.\footnote{Please note that this value is not equal to the threshold at 177\,keV (section \ref{subsec:chkv-effect}) above which the Cherenkov effect can occur based on equation \ref{eq:cos-beta}, that is, when $\beta \cdot n > 1$.}

\begin{figure}
\begin{center}
\includegraphics[width=0.8\textwidth]{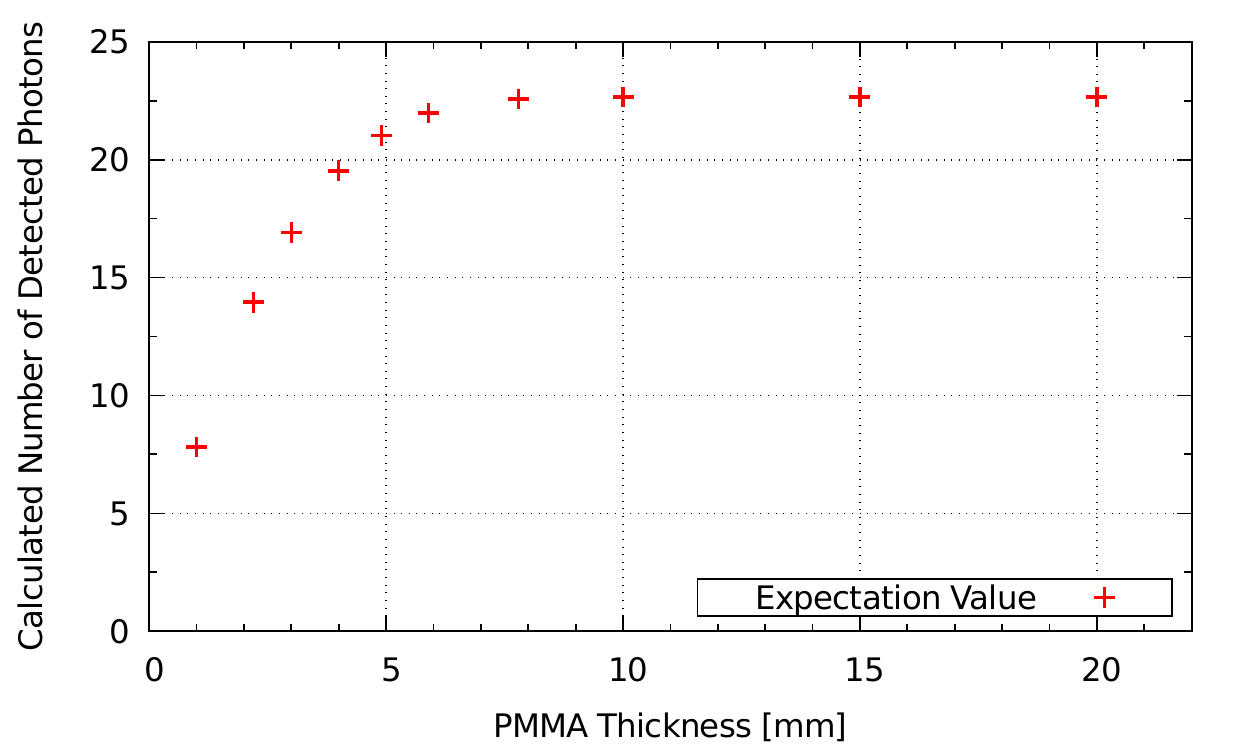}
\caption{Expectation value for the number of detected Cherenkov photons based on the electron energies from a ${}^{90}$Sr spectrum and taking into account PMMA transmission and bias-dependent detection efficiency of the SiPM.}
\label{fig-expected-numPh}
\end{center}
\end{figure}

In the experiment a ${}^{90}$Sr electron source which has a continuous energy spectrum up to 2.4\,MeV is used. 
The goal is to calculate an expectation value for the average number of detected Cherenkov photons per measurement.
Thus, both the distribution of electron energies from the ${}^{90}$Sr source as well as the thickness of the PMMA sample have to be taken into account.
The starting electron energies were chosen from the ${}^{90}$Sr spectrum and the calculation was performed as explained in section \ref{subsec:num-gen-ph}. 
Figure \ref{fig-expected-numPh} shows the results from this calculation. 
The starting energy of the electron depends on the spectrum which therefore also determines the number of created Cherenkov photons.
For smaller thickness, the range of electrons with higher energies is limited by the thickness of the sample and the electron therefore leaves the PMMA and cannot develop its full light yield. 
Above a certain thickness all electrons reach their full range inside the PMMA, and the number of detected photons saturates. 

The size of the Cherenkov cone intersecting the optical readout plane increases with increasing PMMA thickness. 
However, the limited geometrical boundaries of the SiPM array were not taken into account in these estimations. 
Therefore, in the experiment the number of detected photons is expected to drop again as soon as the diameter of the Cherenkov cone intersecting the SiPM's surface is larger than the side length of the SiPM array. This behavior is investigated in the measurements as well (see section \ref{sec:measurements}). The PMMA sample was chosen to be large enough so that there are no reflections from the side walls of the sample.

\section{Experimental Set-Up and Materials}
\label{sec:setup}

\begin{figure*}
\begin{center}
\includegraphics[width=1.0\textwidth]{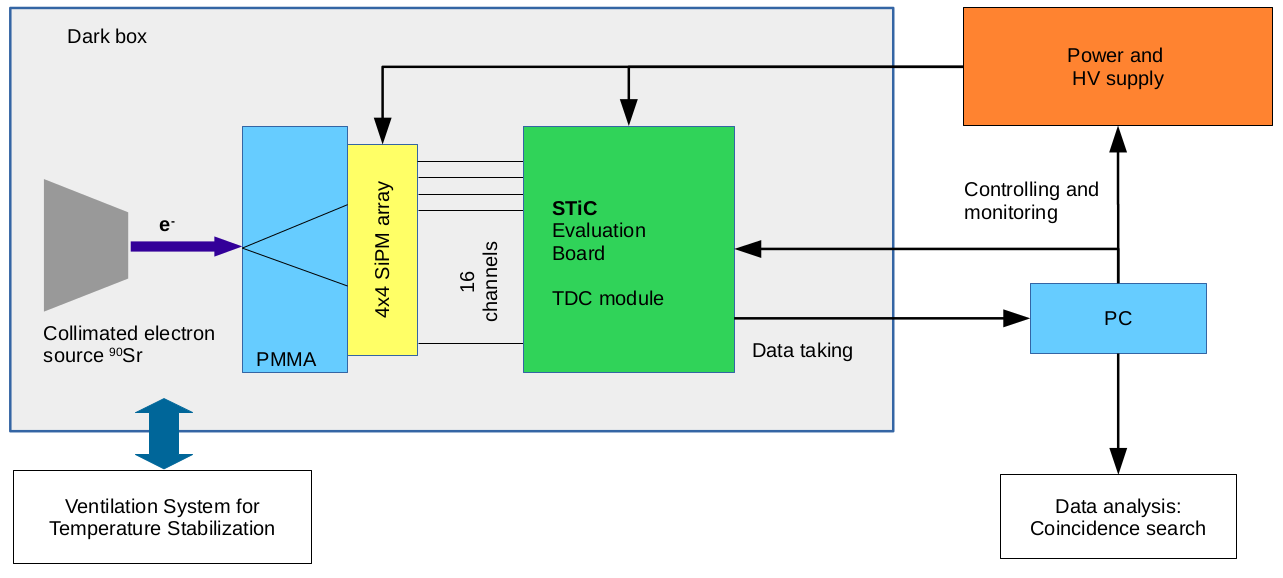}
\caption{Schematic drawing of the set-up for coincident Cherenkov light detection. The whole set-up was placed inside a ventilated dark box to prevent external light pollution.}
\label{fig-set-up-schematics}
\end{center}
\end{figure*}

The measurement set-up consists of a SiPM array with 4x4 channels from Hamamatsu (S13360 series)\cite{ref:hamamatsu} with a pixel pitch of 75\,$\mu$m and a channel size of 3x3\,$\mathrm{mm}^2$. 
The large pixel pitch results in a fill factor of 82\%, which significantly increases the photon detection efficient (PDE).
A coating with silicone resin was chosen to improve the sensitivity in the near UV range. 
PMMA samples, with a density of $\unitfrac[1.09]{g}{cm^3}$,  were optically coupled to the SiPM's surface, using silicone grease in between to match the refractive index of the detector's surface. 
For the PMMA a size of $30\times 30\,\mathrm{mm}^2$ was chosen, which is significantly larger than the SiPM area to avoid reflections of light from the side walls of the sample. The mean refractive index of PMMA in the visible wavelength range is 1.49 \cite{ref:sultanova}. Samples with a thickness of 2.2\,mm, 3.0\,mm, 4.0\,mm, 4.9\,mm, 5.9\,mm, 7.8\,mm, 10\,mm and 15\,mm were used for the experiments.
The sample was irradiated using electrons from a ${}^{90}$Sr source with a maximum energy of 2.42\,MeV. 
The source was collimated to a spot size of 1\,mm in diameter to have a precise positioning with respect to the individual channels of the SiPM. 
The activity of the source at the exit point was calculated to be about 1\,kBq. 

To investigate the influence of transmission and refractive index on the outcome of the measurements, another material called TPX - RT18 \cite{ref:tpx} was used, which is a type of Polymethyl Pentene. A 6\,mm thick sample was available with a density of $\unitfrac[0.833]{g}{cm^3}$ and a refractive index of 1.46. TPX has a significantly larger transmission in the range between 290\,nm and 360\,nm than PMMA, therefore, a larger number of detected photons is expected.

For the purpose of comparison of the distribution of Cherenkov light to scintillation light, measurements were also made using a 9.8\,mm thick sample of Poly-vinyl-tuolene (PVT) with a density of $\unitfrac[1.032]{g}{cm^3}$. This fast scintillator has a decay time of 2.1\,ns and was coupled in the same way as PMMA. 


The readout system uses a SiPM timing chip (STiC) which can read out up to 64 channels simultaneously\cite{ref:stic2014}. 
A threshold can be set on the amplifier output of each channel individually.
The time stamp and time-over-threshold (ToT) value of a channel were recorded whenever the output of the SiPM exceeded the threshold for this channel.
The STiC and its readout evaluation board uses a two trigger method giving each triggered channel a time stamp with 50.2\,ps step size and a ToT value with 1.6\,ns step size, the latter of which is related to the number of photons detected by that channel.
Data analysis and coincidence search was performed offline (for details see section \ref{subsec:data-taking-and-analysis}).

To prohibit light pollution from external light sources the measurements took place in a dark box. A ventilation system kept stable conditions at room temperature. The whole set-up is shown schematically in figure \ref{fig-set-up-schematics}.

To compare and confirm the results obtained with the STiC-set-up, a second measurement technique was performed using a 4-channel mixed signal oscilloscope with a 4\,GHz bandwidth. 
A fixed SiPM channel was used as trigger while looking for coincident signals on the other three channels. Charge integration of the SiPM signal allowed for counting of photons of each coincident event in a channel.
For this measurement the signal of the SiPM was amplified by a factor of 50 using an external, fast amplifier with a rise time of 220\,ps.

\section{Measurements}
\label{sec:measurements}
\subsection{Data Taking and Analysis}
\label{subsec:data-taking-and-analysis}

For the measurements in this section the SiPM was operated at constant room temperature with an overvoltage of 4.2\,V,  while for the photon counting measurements in section \ref{subsec:count-chkv-photons} a voltage of 3.8\,V was set.
The signal of any channel was recorded when it was above the trigger threshold for this channel.
Coincidence filtering during the measurement was not performed. In order to reduce the contribution of dark count the thresholds were set well above the signal level of one photon. 
On a 2-3 pe level (pe = photon-equivalent) the dark count rate was in the range of 1-10\,kHz per channel.
For the purpose of achieving best coincidence timing resolution, a time calibration was performed to compensate for time difference between individual channels inherent in the ASIC of the STiC. 

Data was taken over a period of 60 seconds for each PMMA sample with a thickness between 2.2\,mm and 15\,mm. The analysis was performed offline and the signals from all channels were sorted by their time stamp.
Starting with the earliest signal, all channels within a time window of 800\,ps were combined into one \textit{coincident event} if there were at least three channels triggered within that time window.

\begin{figure}
\begin{center}
\hspace*{-0.3cm}
\includegraphics[width=0.8\textwidth]{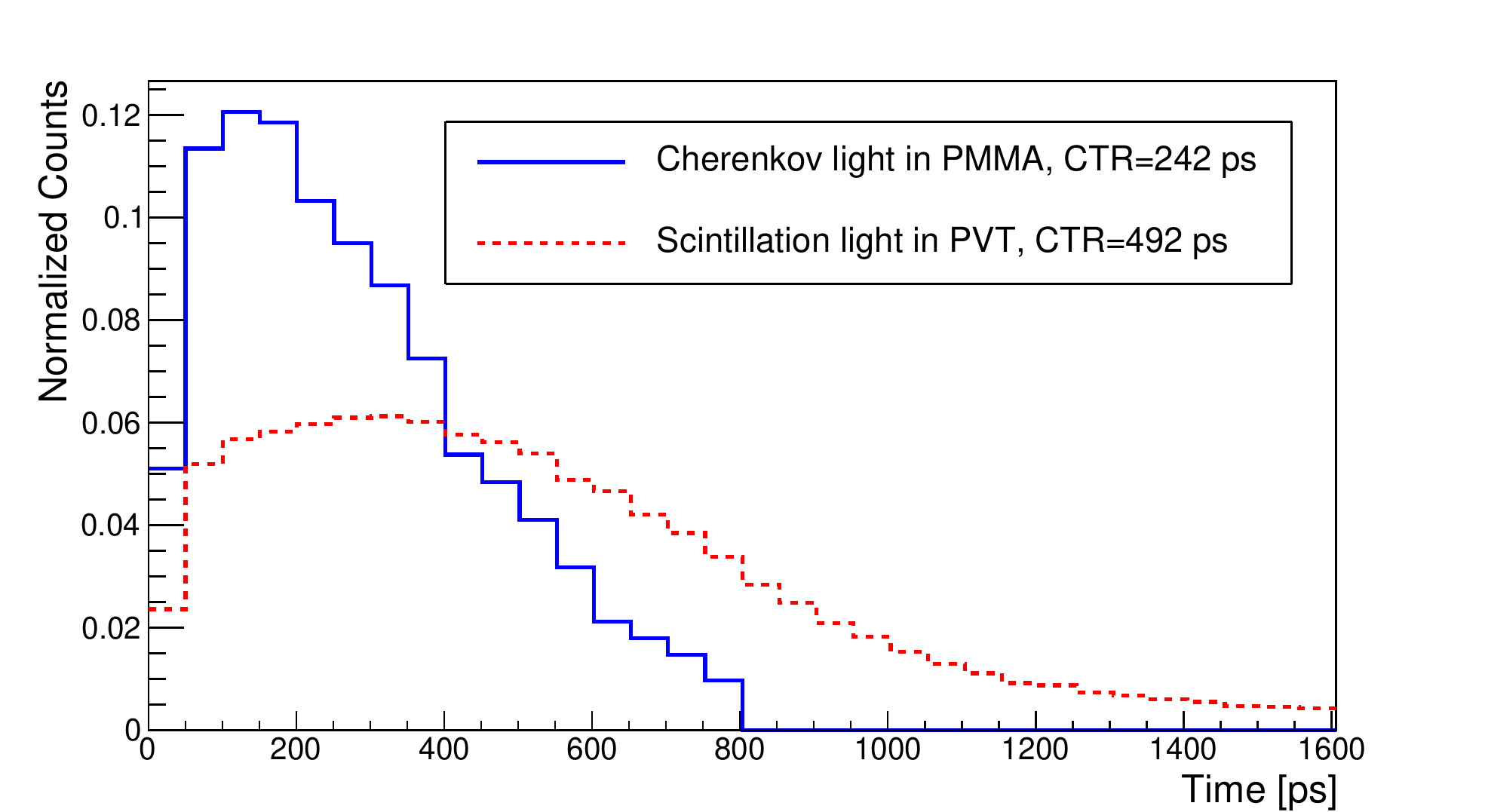}
\caption{Coincidence time of Cherenkov light (solid line) from electrons in PMMA and scintillation light (red dotted line) created in a fast scintillator PVT. Cherenkov light is emitted instantaneously while scintillation light follows an exponential behavior and therefore shows a longer tail.}
\label{fig-CTR}
\end{center}
\end{figure}

The coincidence time resolution (CTR) was determined using the temporal distance between each coincident channel with respect to the first one involved in the event. 
Figure \ref{fig-CTR} shows the temporal distance of the triggered channels for a 10\,mm sample size. 
The mean value of this distribution is at 242\,ps, which is in good agreement with the timing resolution of 180\,ps obtained in characterization measurements for the STiC\cite{ref:stic2014}.
It is worth mentioning that the measured CTR was independent of the thickness of the sample. 
This measurement was compared to the coincidences using electrons in the fast scintillator PVT. 
As shown in Figure \ref{fig-CTR}, the scintillation light exhibited a larger temporal spread than that of Cherenkov light, with a mean value of 492 ns.\footnote{For a better representation of this behavior, the coincidence search window has been increased from 800\,ps to 1.6\,ns in the graph of figure \ref{fig-CTR}.}
This demonstrates the instantaneous nature of the creation of Cherenkov light in distinction to scintillation light.

\subsection{Distribution of Coincident Cherenkov Photons}
\begin{figure}
\begin{center}
\includegraphics[scale=0.14]{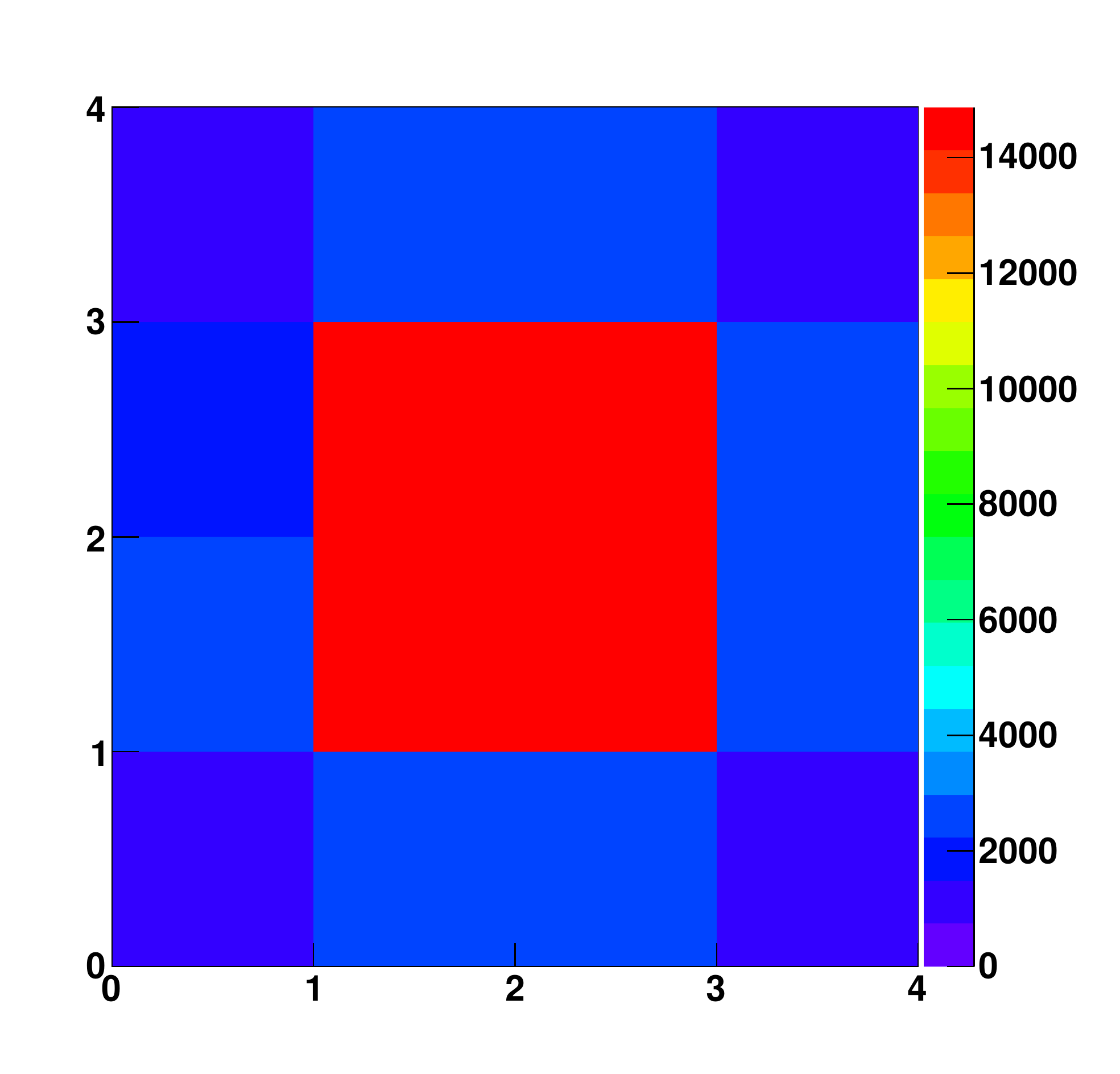}
\includegraphics[scale=0.14]{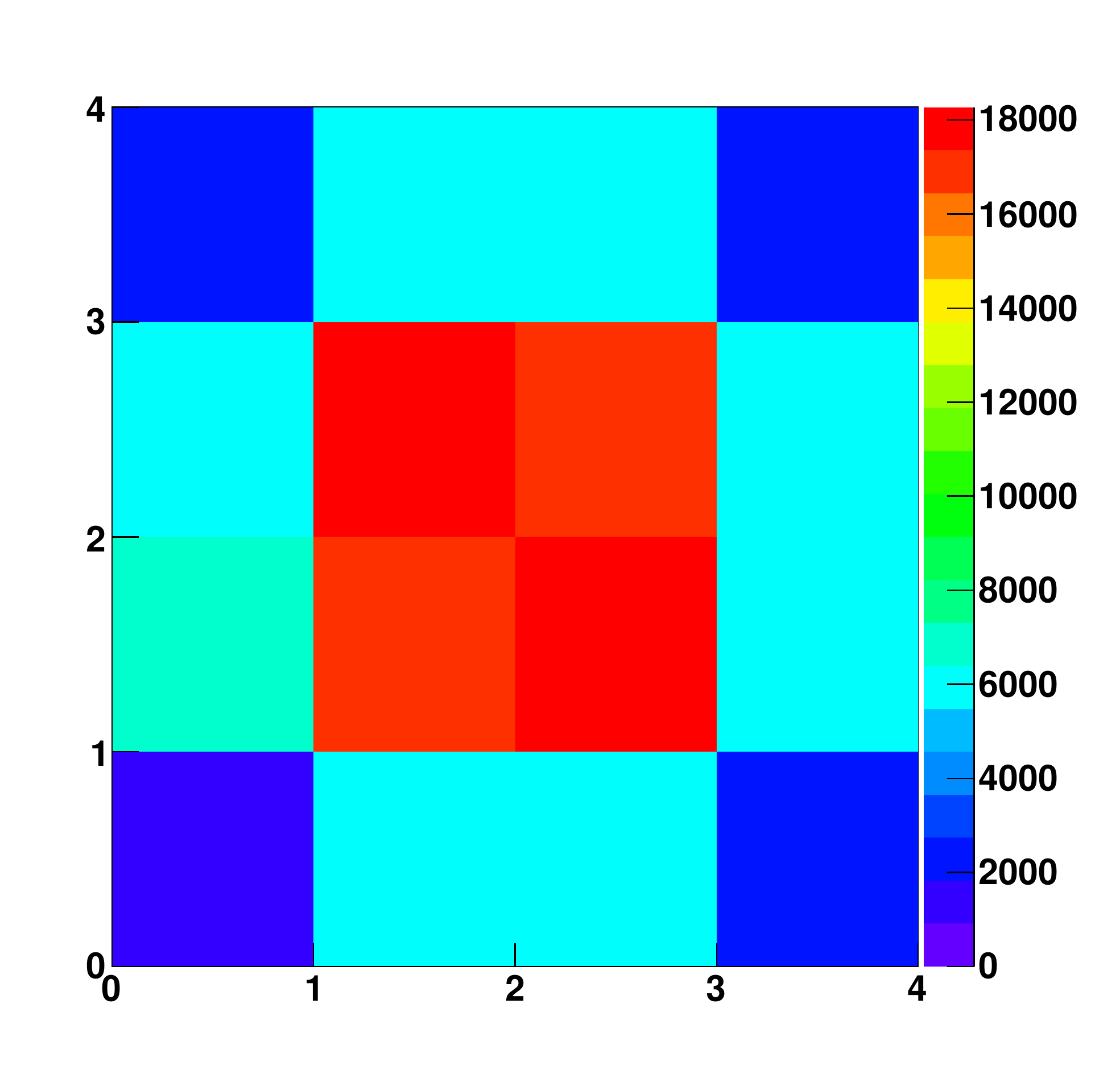}
\includegraphics[scale=0.14]{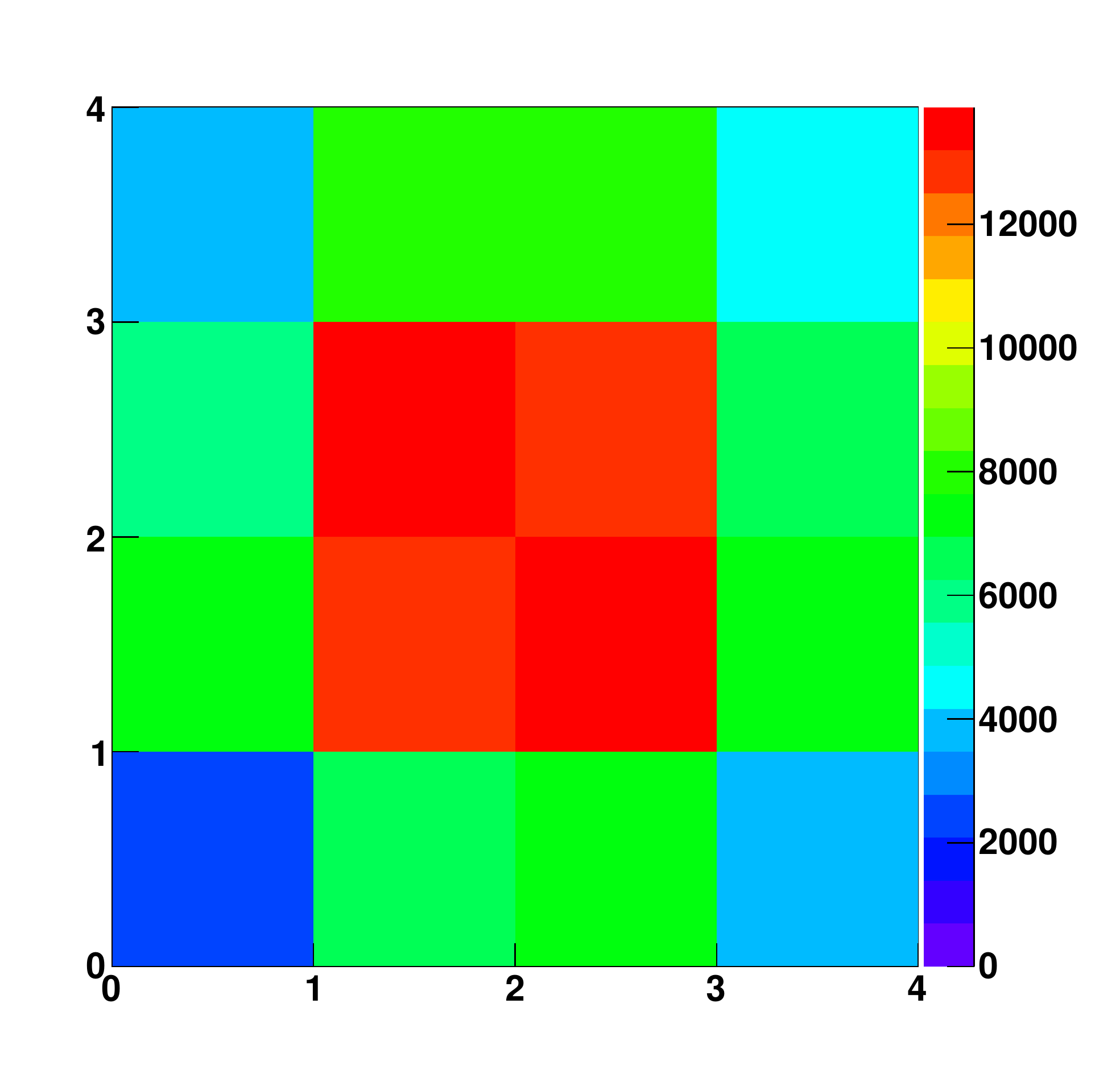}
\includegraphics[scale=0.14]{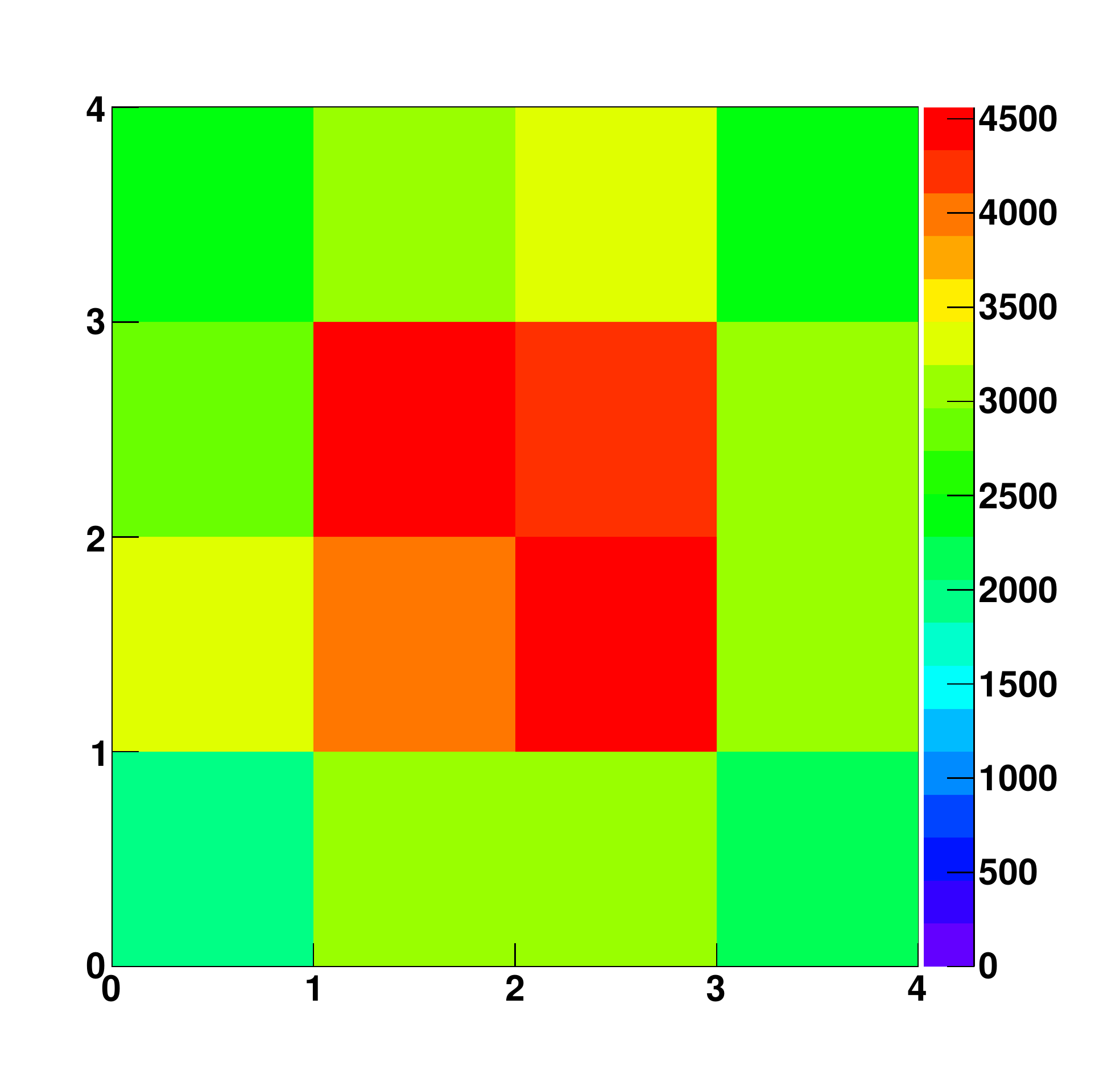}
\caption{Occupancy of the $4\times4$ SiPM array for various PMMA thicknesses. Shown is the distribution of coincident channels with the electron source centered above the array. Thickness of the PMMA sample from top left to bottom right: 2.2\,mm, 4.0\,mm, 5.9\,mm and 10.0\,mm. The Cherenkov cone radius increases with sample thickness.}
\label{fig-sipm-center}
\end{center}
\end{figure}

Measurements were performed for different thicknesses of PMMA varying from 2.2\,mm to 15\,mm and for a TPX sample with a thickness of 6\,mm. All channels from each coincident event were recorded. The electron source was pointed at the center of the array so that a symmetric Cherenkov cone in the middle of the SiPM array is expected. Figure \ref{fig-sipm-center} shows the distribution of the coincident events over the array for various thicknesses: The colors indicate the number of occurrences of each channel in a coincident event. As expected, the width of the distribution increases with larger thickness and the relative difference between the center channels and the outer ones decreases. This observation represents the expanding cone radius of the Cherenkov light with increasing thickness. While in a 2.2\,mm sample only the channels in the center are involved in coincident detections, for greater thicknesses also the outer channels are triggered. 

\begin{figure}
\begin{center}
\includegraphics[scale=0.14]{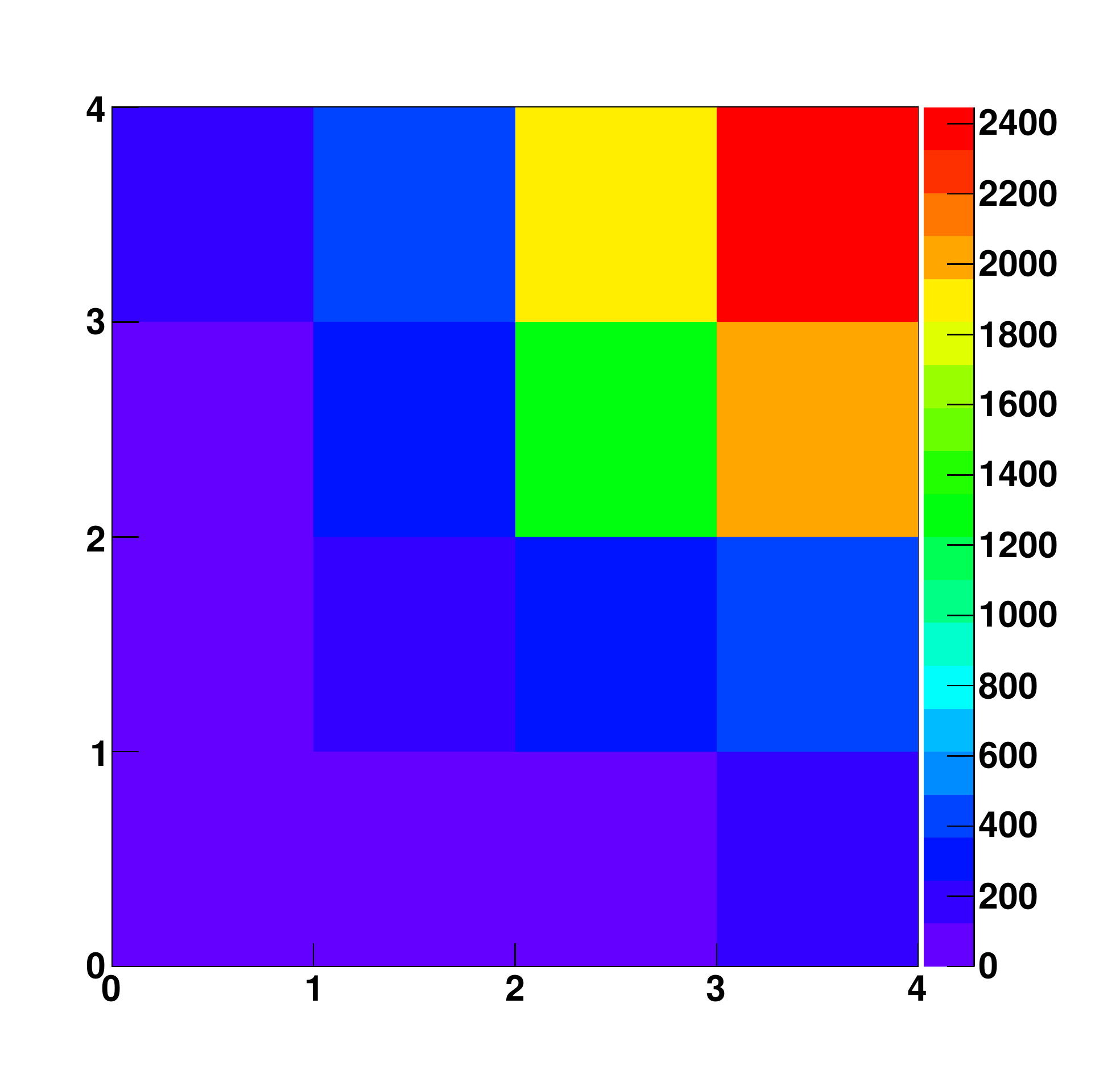}
\includegraphics[scale=0.14]{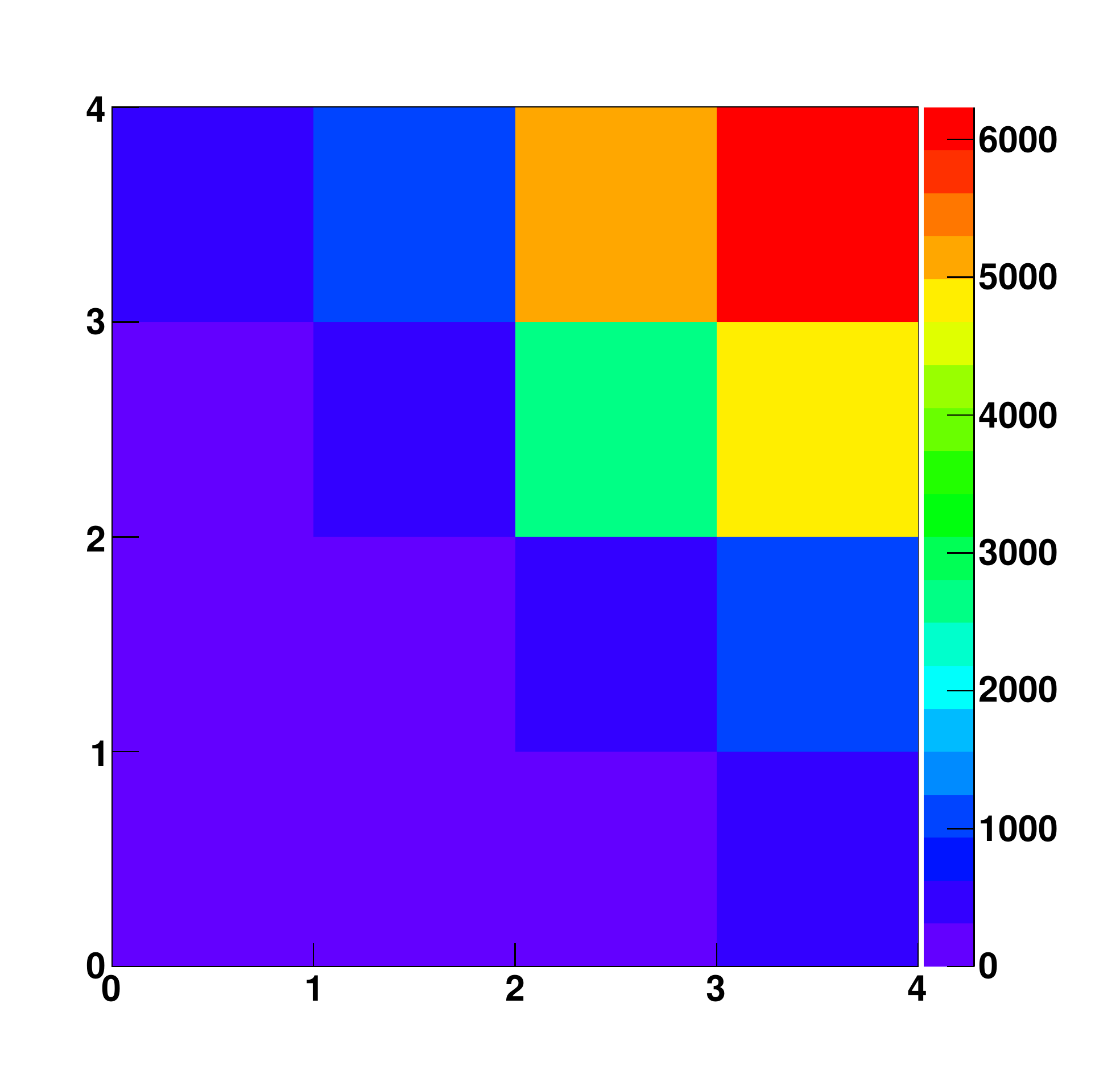}
\includegraphics[scale=0.14]{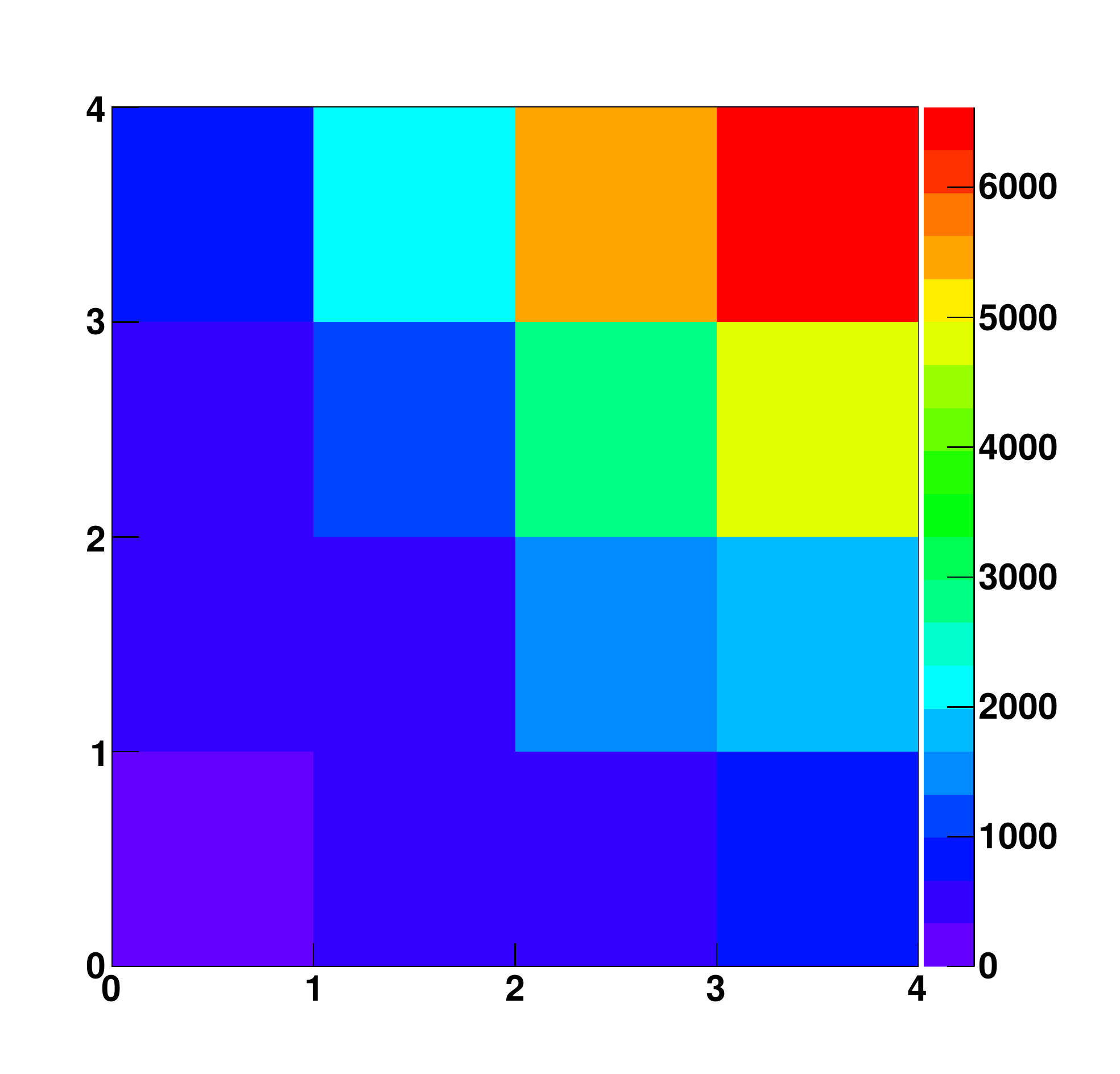}
\includegraphics[scale=0.14]{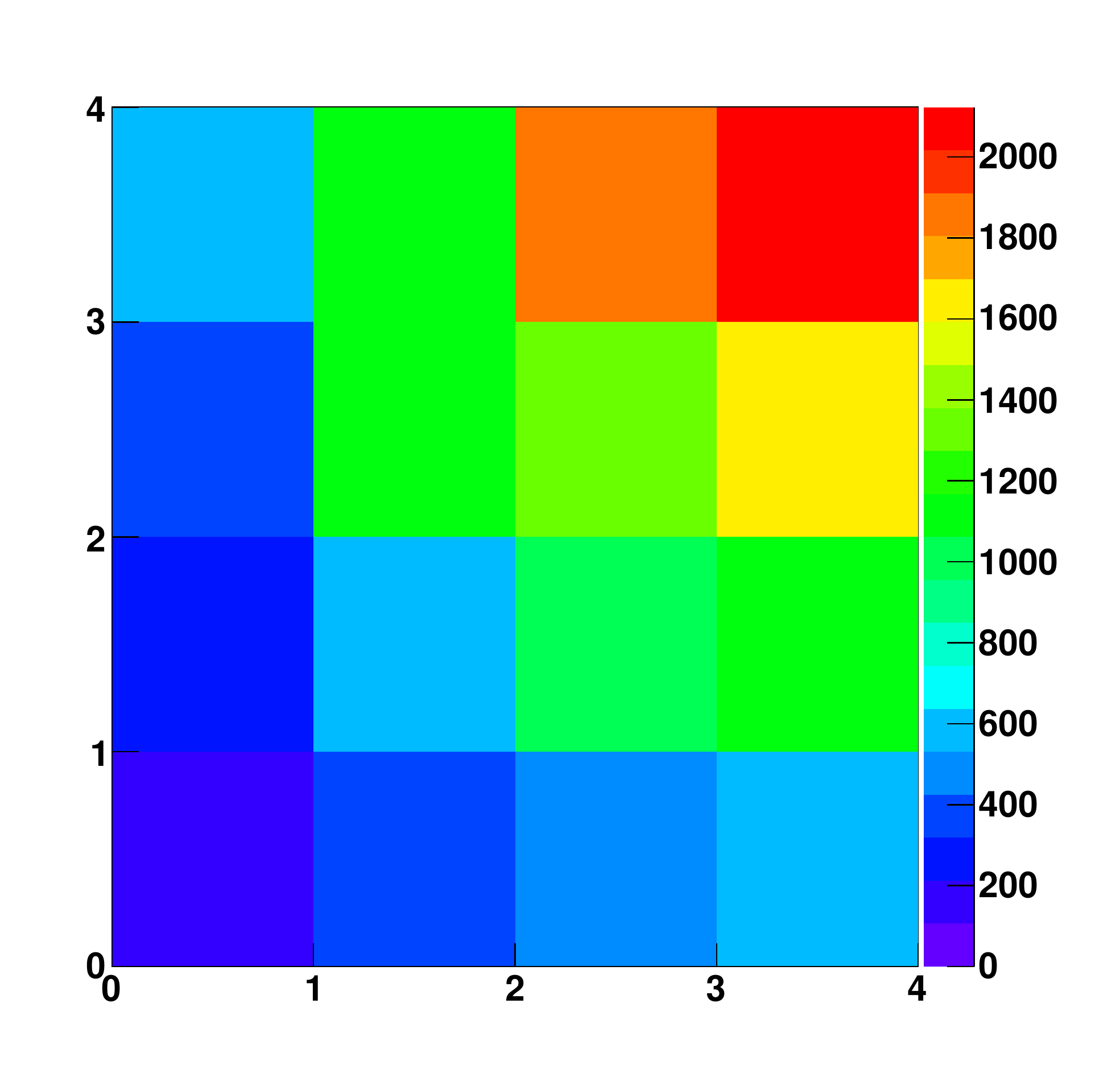}
\caption{Occupancy of the $4\times4$ SiPM array for various PMMA thicknesses. Shown is the distribution of coincident Cherenkov photons created in PMMA with the electron source pointed at the corner of the array. Thickness from top left to bottom right: 2.2\,mm, 4.0\,mm, 5.9\,mm and 10.0\,mm.}
\label{fig-sipm-corner}
\end{center}
\end{figure}

In a second measurement the source was moved to the top right corner of the array. Again the occupancy of the SiPM array was examined. Results are shown in figure \ref{fig-sipm-corner}. For a quantification of the spatial distribution of Cherenkov photons see section \ref{sec:results}.

\begin{figure}
\begin{center}
\includegraphics[width=0.45\textwidth]{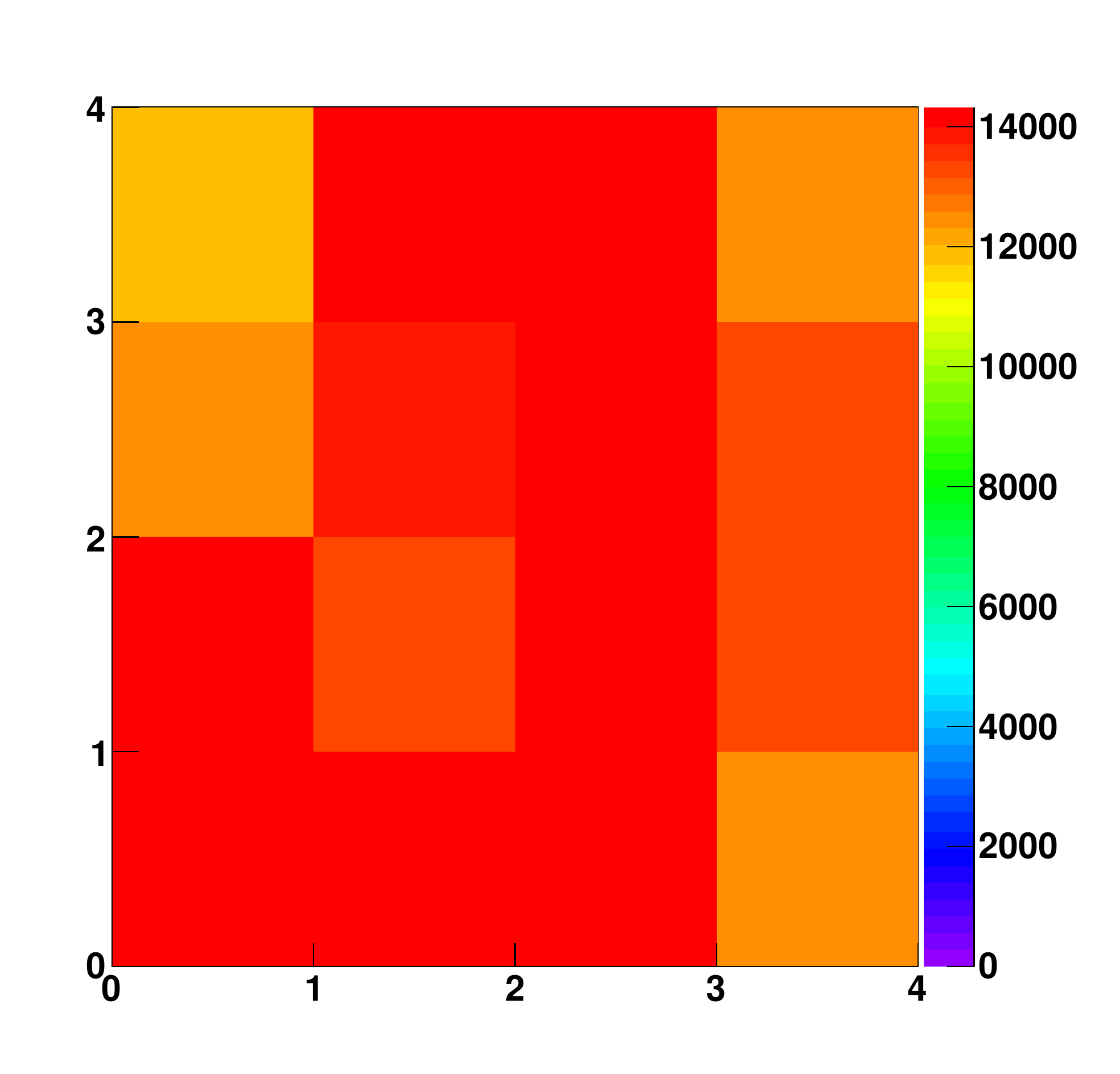}
\caption{Occupancy of the SiPM array using scintillation light. Unlike Cherenkov light, scintillation light triggers all channels with nearly equal frequency, creating an even distribution on the detector array.}
\label{fig-scintillator}
\end{center}
\end{figure}

A further and more significant distinction between scintillation and Che\-renkov light can be made by looking at the distribution of coincident hits on the array. While Cherenkov light  follows the characteristic cone with its corresponding energy-dependent opening angle, scintillation light is emitted isotropically all along the electron track and emitted photons should trigger all channels. This behavior could be shown by replacing the PMMA sample with the aforementioned scintillator PVT with a thickness of 9.8\,mm. Again, the entire detector array was covered by the scintillator. The isotropic nature of the emission of scintillation light can be seen in the occupancy plot in figure \ref{fig-scintillator} . 

\subsection{Quantifying the Light Intensity using Time over Threshold Measurements}
\label{subsec:quantify-light-yield}
The time over threshold (ToT) signals from the STiC were used to estimate the energy information of each event. The ToT is closely related to the height of the signal created in a channel, which again is proportional to the number of photons detected by that same channel. Therefore, this information can be used as a measure of the intensity of the detected Cherenkov radiation. In the present case, the bin size of the time over threshold measurement was 1.6\,ns, which was sufficiently small considering a SiPM peak width on the order of 100\,ns. For the purpose of measuring the intensity, the energy information from each coincident channel per event was added up. The mean value of these energy sums for all coincident events was used to quantify the detected light level in arbitrary units. A repetition of this procedure for all sample thicknesses enabled the drawing of conclusions on the change in number of photons with thickness of the PMMA sample. 
Since light detection on single photon level was not possible with this device, a calibration of the energy to a number of photons was not performed and arbitrary units are used.
\subsection{Counting Coincident Cherenkov Photons}
\label{subsec:count-chkv-photons}

As mentioned in section \ref{sec:setup}, a second set-up using an oscilloscope was used that enables photon counting in each channel. For that purpose, the charge information and its proportionality to the number of detected photons in each channel was used. Integrating the measured voltage signal over a fixed time interval $[t_0,\,t_1]$ allows for a calculation of the deposited charge as follows:
\begin{equation}
Q = \frac{1}{R\cdot G}\int_{t_0}^{t_1} V(t)\,\mathrm{d}t
\end{equation}
where $R$ denotes the resistance creating the voltage drop and $G$ is the gain. 

For this measurement the overvoltage was 3.8\,V. The voltage signal of a channel was amplified with a gain of 50 and fed into a 4\,GHz mixed signal oscilloscope. An integration time of 200\,ns was chosen, which was long enough to capture the full length of the signal and still have very small probability for a random dark count event within that time interval. The measurement procedure involved fixing one channel in the center of the array as trigger while three others were read out. Therefore, any measured light signal was in coincidence with the trigger channel. The measured signals were corrected for base line fluctuations and the calculated charge values were used to create a peak integral spectrum with 10000 triggered events. 
The discrete nature of the peaks in the spectrum enables to calibrate each channel and thereby to assign a number of detected photons to each measured charge value.
 This allows for counting the number of photons that were measured in that channel for each triggered event. Following this procedure all 16 channels were read out and analyzed. In the last run, a different trigger channel was chosen and the former trigger channel was read out in the same way. 

The average number of detected photons in each channel was calculated as well as the total sum from all channels. The latter is used to compare measurements with different sample thicknesses with one another. A measurement without an electron source was used to estimate contributions from dark counts and, subsequently, subtract it from the obtained number of photons.
The trigger level has a visible influence on the result as it cuts into the energy spectrum of the electrons from the ${}^{90}$Sr source. 
\FloatBarrier
\section{Results}
\label{sec:results}

\subsection{Cherenkov Light Intensity and Number of Photons}
\label{subsec:intensity}
\begin{figure}
\begin{center}
\includegraphics[width=0.8\textwidth]{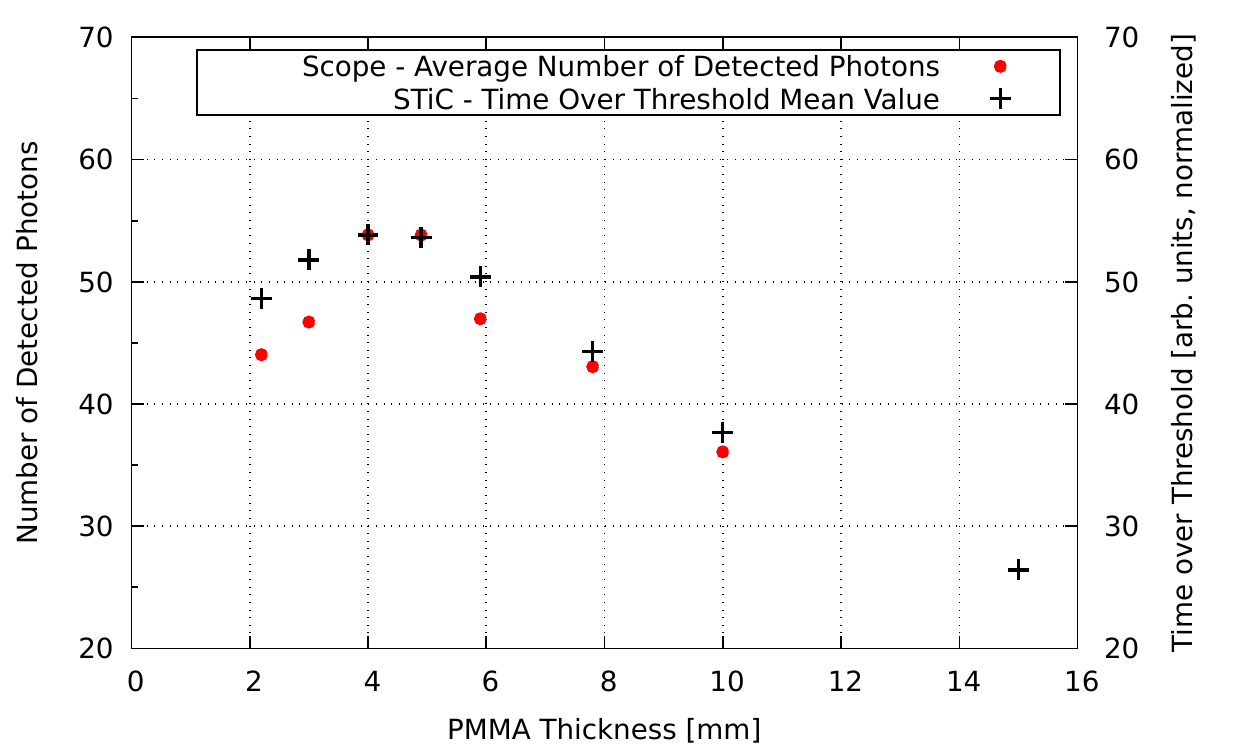}
\caption{For various thicknesses: Counted number of coincident photons from the charge integration using the oscilloscope (circles) and measured light intensity using the ToT information from the STiC set-up (pluses). The STiC results have been normalized to the 4.0\,mm measurement from the scope.}
\label{fig-light-intensity-stic-scope}
\end{center}
\end{figure}

The measurements described in section \ref{subsec:count-chkv-photons} were conducted for PMMA thicknesses between 2.2\,mm and 10\,mm. The results are depicted in figure \ref{fig-light-intensity-stic-scope} together with results from STiC measurements.
A list of all results can be found in table \ref{tab-results}.

At first, the average number of detected photons increases with increasing thickness forming a plateau-like structure at 4-5\,mm before a decline occures above 5.9\,mm. 
In the lower range the dominating effect is the limited electron range determined by the thickness of the PMMA, resulting in a reduction of the Cherenkov light yield. 
For the higher energetic fraction of electrons, this causes the Cherenkov effect to terminate before the Cherenkov energy threshold is reached and, therefore, reduces the light yield. This reduction decreases with growing thickness explaining the increasing number of detected photons between 2.2\,mm and 4\,mm.
In PMMA the maximum possible Cherenkov cone opening angle is $47.8^\circ$. 
This means with higher thickness the radius of the cone intersecting the surface of the SiPM is larger than the width of the array (the SiPM has a side length of 12\,mm, compare section \ref{subsec:count-chkv-photons}). This implies that above a thickness of 5.7\,mm for the PMMA absorber the number of detected photons starts to decrease again due to the Cherenkov cone exceeding the boundaries of the SiPM. The same behavior can be reproduced by the STiC's time over threshold measurements. Since the ToT value is closely related to the number of detected photons, it can be used to quantify the incident light level. As figure \ref{fig-light-intensity-stic-scope} indicates, the value increases for smaller thicknesses and starts dropping again for all samples thicker than about 5.7\,mm.

The ToT measurement was also performed for a sample of TPX with a thickness of 6\,mm. The refractive index, which is a key quantity for the number of created Cherenkov photons according to formula \ref{eq:cos-beta}, is 1.46 and therefore slightly smaller than the one of PMMA. Nevertheless, an increase in the detected light intensity is expected due to the higher transmission in the near UV range below 400\,nm \cite{ref:tpx}. The ToT measurement for this TPX sample gives a light level which is 30.1\,\% higher compared to the one from the 5.9\,mm sample of PMMA. This denotes a significant increase within the uncertainties of our measurements. Calculations predict an average of 37.7 detected photons for this TPX sample compared to 22.0 photons for the 5.9\,mm sample of PMMA.

\begin{table}[h]
\begin{center}
\begin{tabular}{@{}cccc@{}}
\toprule
Material & Thickness & \begin{tabular}{@{}c@{}}Number of\\Detected Photons\end{tabular} & \begin{tabular}{@{}c@{}} Time over\\Threshold\end{tabular} \\ \midrule
		PMMA & 2.2$\,$mm & 44.0$\,\pm\,$0.4 & 48.6$\,\pm\,$0.3 \\
         &3.0$\,$mm & 46.7$\,\pm\,$0.5 & 51.8$\,\pm\,$0.3\\
         &4.0$\,$mm & 53.8$\,\pm\,$0.5 & 53.8$\,\pm\,$0.3\\
         &4.9$\,$mm & 53.8$\,\pm\,$0.5 & 53.7$\,\pm\,$0.3\\
         &5.9$\,$mm & 47.0$\,\pm\,$0.5 & 50.4$\,\pm\,$0.3\\
         &7.8$\,$mm & 43.1$\,\pm\,$0.4 & 44.3$\,\pm\,$0.3\\
         &10.0$\,$mm & 36.1$\,\pm\,$0.4 & 37.7$\,\pm\,$0.4\\
         &15.0$\,$mm & -- & 26.5$\,\pm\,$0.7\\ \midrule
         TPX & 6\,mm & -- & 65.6$\,\pm\,$0.4\\ \bottomrule
\end{tabular}
\caption{Results for the detected number of Cherenkov photons from electrons in PMMA and TPX measured with the oscilloscope set-up together with the corresponding Time over Threshold values.}
\label{tab-results}
\end{center}
\end{table}

\subsection{Comparison to Theoretical Expectation}
\label{subsec:comparison}

\begin{figure}
\begin{center}
\includegraphics[width=0.8\textwidth]{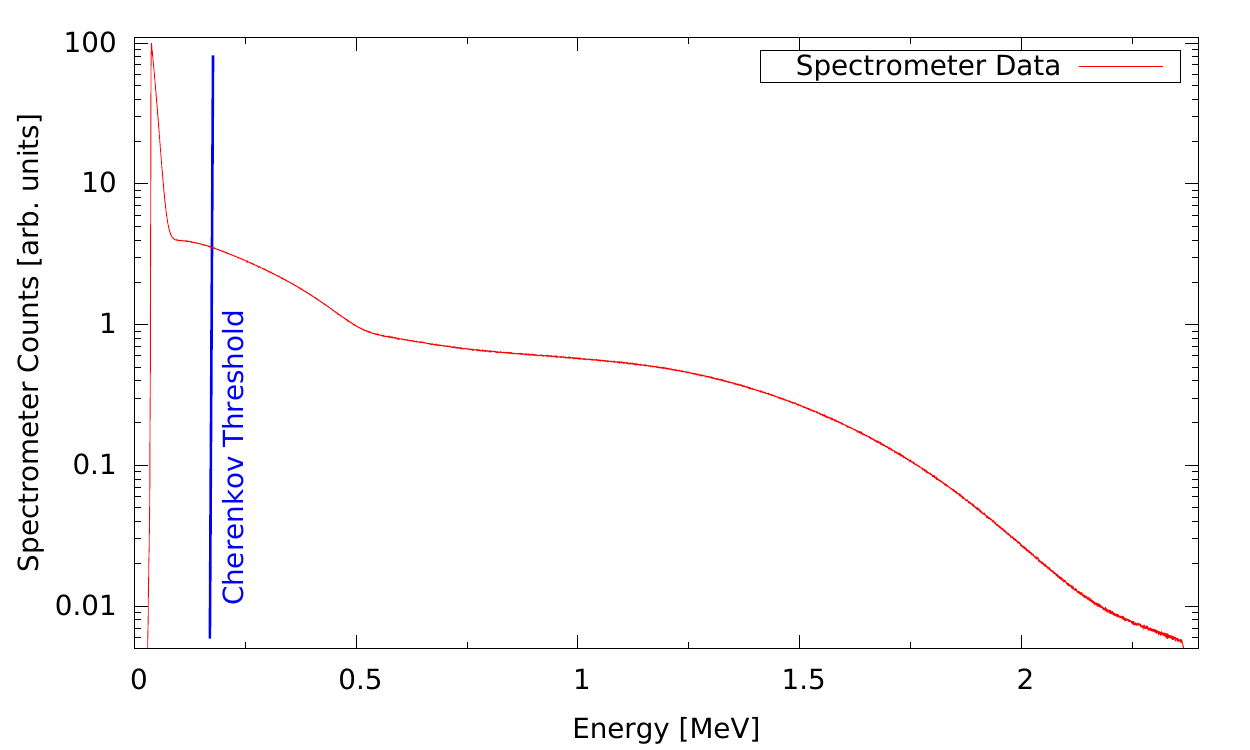}
\caption{Strontium energy spectrum: Especially lower energies show higher contributions to the spectrum.}
\label{fig-strontium-spec}
\end{center}
\end{figure}

A comparison of above presented results for the number of detected photons with the scope measurement with the predictions made in chapter \ref{sec:theory} reveals the same qualitative behavior for PMMA thicknesses up to 5.9\,mm: 
In that range the number of photons increases with growing thickness.
One finds that the calculations predict a growth of 13.4\,\% between 3\,mm and 4\,mm. In the measurement a gain of 13.2\,\% is obtained in the same range. 
The calculations were performed using a minimal energy of 201\,keV, the energy required to produce at least one photon in the investigated wavelength range. 
Absolute numbers, however, deviate significantly: While only 19.5 detected photons were calculated for a thickness of 4\,mm, in the actual measurement 53.8 photons were detected. 
The most important contribution to this deviation lies in the threshold of the trigger channel.
The trigger threshold was set above the level of 2 photons, in order to reduce the contribution of dark signals to the measurement.
At this level the dark count rate is reduced by about two orders of magnitude. 
The trigger threshold induces a significant bias on the detection: 
Lower energetic electrons, producing a small number of photons, are less likely to be detected. The electron energy spectrum of $^{90}$Sr in figure~\ref{fig-strontium-spec} shows a strongly increased intensity towards smaller electron energies where fewer photons are generated. Therefore, in the calculation these energies have a large contribution to the average number of detected photons, while in the measurement these events are most likely not triggered. Calculations show that the number of detected photons increases to a value of 35.1 if only electrons with a kinetic energy of at least 0.5\,MeV are considered. The measured value of 53.8 is reproduced by the calculation when using a minimum required electron energy of 1.4\,MeV. This confirms the influence of the trigger level on the number of detected Cherenkov photons. 

Unlike in the theoretical estimation where it stays flat, in the measurement above 5.9\,mm the detected photon number starts to drop. In the described measurement set-up some fraction of the photons will be outside the active area of the detector due to the limited area of the SiPM. This is one reason why in this set-up the number of detected photons depends on the thickness of the material. 

Another reason for the discrepancy is multiple scattering of the electrons, which effectively enlarges the total range of the electron until being stopped by the SiPM's surface. Since according to equation \ref{eq:N_i} the number of emitted photons depends on the travelled distance of the electron, the number of detected photons increases as well. Lower electron energies in the range of up to 1.5\,MeV would undergo that effect and especially these energies hold the major contribution to the Sr spectrum. However, the number of photons created in that energy range is lower than for higher energies. 
Furthermore, as the electron scatters inside the radiator, some fraction of Cherenkov photons will not reach the active area of the detector.
The quantification of the contribution of multiple scattering is therefore subject to further investigations.



\begin{figure}
\begin{center}
\includegraphics[width=0.8\textwidth]{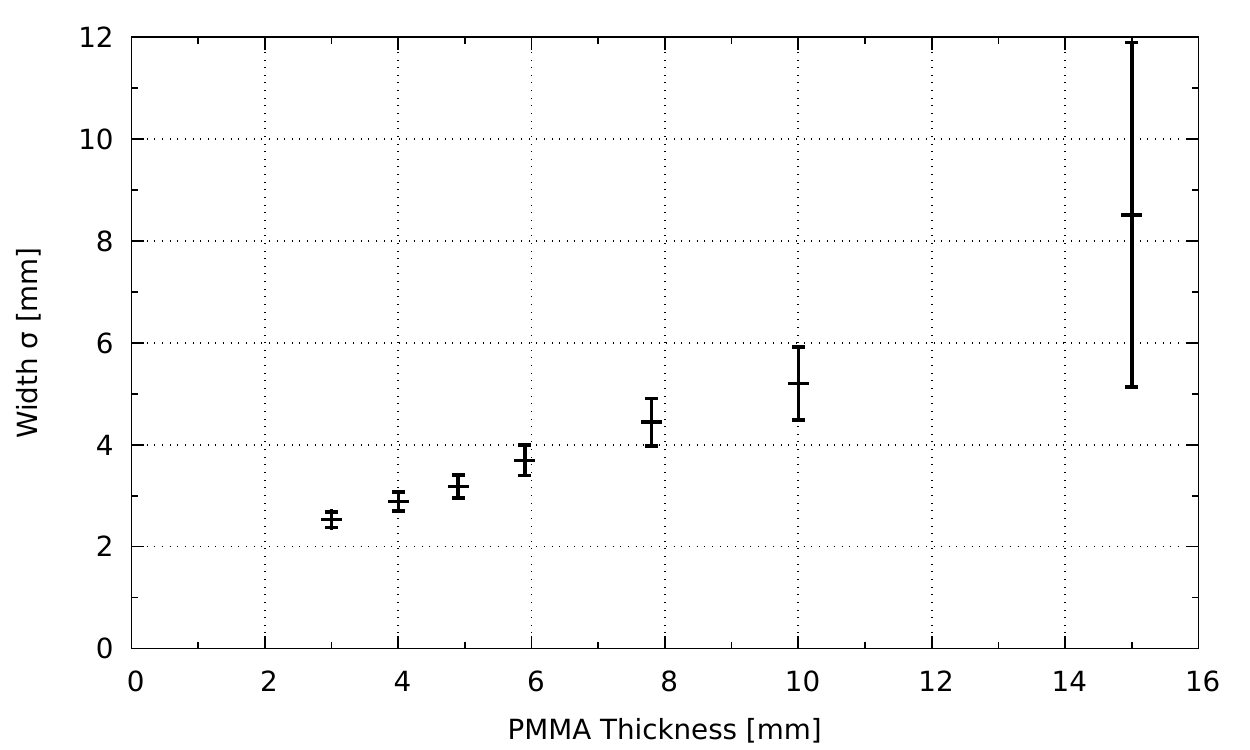}
\caption{For various thicknesses: Spread of the distribution of coincident Cherenkov photons quantified by the width of a gauss fit.}
\label{fig-width}
\end{center}
\end{figure}

\subsection{Quantifying the Occupancy of the SiPM Array}
\label{subsec:gauss-width}
As mentioned in section \ref{subsec:intensity} the number of detected Cherenkov photons depends on the thickness of the PMMA radiator. 
This is mostly due to the limited active area of the detector. This means for a large enough SiPM array, the number of detected photons should be independent of the thickness of the sample (given a full electron range).
The radius of the intersection between the Cherenkov cone and the detector's surface changes with radiator thickness as well. This behavior has already been shown graphically in figure \ref{fig-sipm-center}. This also implies that the width of the distribution of the photons over the array increases with thickness, which is subject to further investigation in this section. 

To quantify the spread of the distribution a two-dimensional gauss fit was performed using the data from the centered electron source position. 
The width of the distribution was then quantified using the fit parameter $\sigma$. 
A symmetric fit was chosen, leaving the $\sigma$-parameter the same in $x$- and $y$-direction. 
Also, since the source position was defined by the set-up the mean value was fixed as well and not used as fit parameter. 
Figure \ref{fig-width} shows the results of this analysis technique for different thicknesses of PMMA.
Especially for a thickness of 15\,mm the coarseness of the sampling (only 16 detector channels) causes enhanced uncertainties.
Nevertheless, the expected behavior for a continuous increase of the width with growing thickness could be reproduced. 

\begin{figure}
\begin{center}
\includegraphics[width=0.5\textwidth]{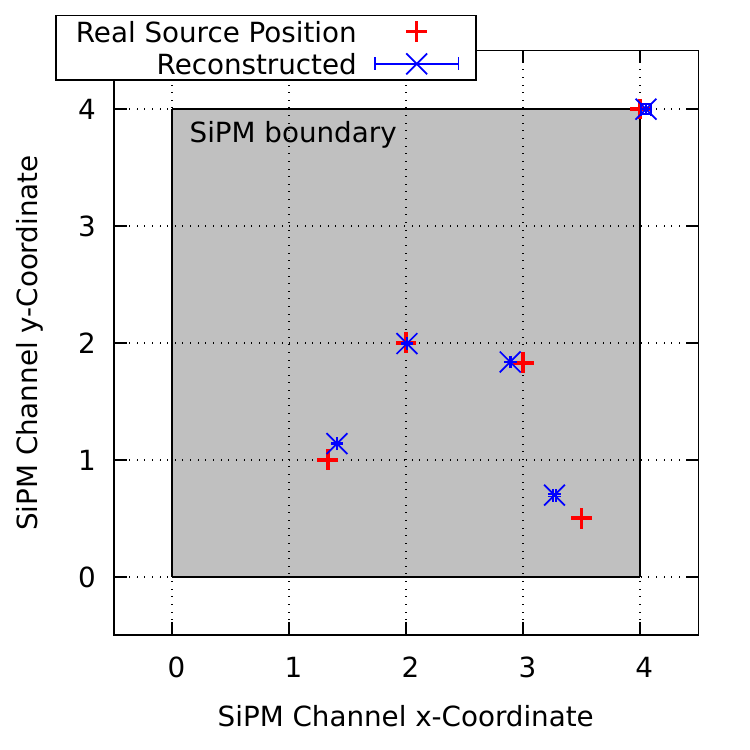}
\caption{Real and reconstructed electron source locations.}
\label{fig-reconstr}
\end{center}
\end{figure}
In comparison to PMMA a sample of TPX with a thickness of 6\,mm was used. While for PMMA (the thickness of 5.9\,mm was chosen for this comparison) the $\sigma$-parameter was $(3.70\pm 0.30)\,$mm, TPX showed a value of $(3.58\pm0.26)\,$mm. 
The difference in refractive index of TPX ($n=1.46$) compared to PMMA ($n=1.49$) causes the opening angle of the Cherenkov cone to be smaller inside the TPX sample according to equation \ref{eq:cos-beta}. In PMMA a maximum opening angle of 47.8$^\circ$ is calculated, while in TPX the angle is 46.8$^\circ$.
Thus, also the $\sigma$-parameter for TPX should be smaller than for PMMA. Within the uncertainties of the measurement no significant difference in the $\sigma$-parameter could be found.

\subsection{Ability to Reconstruct the Electron Source Position}
The electron source was moved towards different positions using a step motor with a position resolution on the scale of $\mu$m. The position was noted and data was taken in the same way as before using STiC and 60\,s of measurement time. A gauss fit was applied to the two-dimensional histogram of the occupancy of the channels with the goal to reconstruct the mean value. 
Unlike in the previous section the mean values in $x$- and $y$-direction were not fixed anymore and rather used for a reconstruction of the various source locations. 
Results are shown in figure \ref{fig-reconstr}. A precise calculation of the source location was possible with an accuracy in the order of 1\,mm, which is mainly influenced by the diameter of the source collimator of 1\,mm.
This result successfully demonstrates a spatial sensitivity for the electron source location using an accumulation of coincident Cherenkov photons.

\FloatBarrier
\section{Conclusion}
\label{sec:conclusion}

In the present paper the coincident detection of Cherenkov photons from high energetic electrons in PMMA was demonstrated using a $4\times 4$ SiPM array. The obtained timing resolution of 242\,ps promises good applicability for medical imaging techniques like a Compton Camera. The electron source location could be reconstructed successfully using the mean value of the distributed Cherenkov photons from many events on the array. In the next steps imaging of single coincident events will be performed, which requires more channels in order to reconstruct the Cherenkov cone on the array. Also, detection on single photon level is envisaged to improve the energy resolution and therefore also the sensitivity for the number of photons that are detected.
This can subsequently be used to reconstruct the momentum direction of single electrons in PMMA. In case of a Compton camera application, the electron would be created by a high-energy gamma scattering in the PMMA sample. Therefore, the electron carries a large part of the momentum information of the incident gamma. This would be the next step towards a working Compton Camera prototype.

 \appendix
 
\vspace{0.5cm}
\section*{Acknowledgements}
We would like to thank the research group around Prof Dr Hans-Christian Schulz-Coulomb for their support with the use of the STiC evaluation kit. The cooperation with Prof Dr Erika Garutti's group from DESY, Hamburg, was very productive and is worthy of special mention at this point.

\vspace{0.5cm}
\section*{Funding}
This research did not receive any specific grant from funding agencies in the public, commercial, or
not-for-profit sectors.


\vspace{0.5cm}
\section*{References}
  \bibliographystyle{unsrt}
  \bibliography{References.bib}

\begin{thebibliography}{10}

\bibitem{ref:Vandenberghe}
S.~Vandenberghe et~al.
\newblock Recent developments in time-of-flight pet.
\newblock {\em EJNMMI Physics}, 3(1):3, Feb 2016.

\bibitem{ref:brunner}
S.~E. Brunner and D.~R. Schaart.
\newblock {BGO} as a hybrid scintillator / cherenkov radiator for
  cost-effective time-of-flight pet.
\newblock {\em Physics in Medicine and Biology}, 62(11):4421, 2017.

\bibitem{ref:peterson}
T.~E. Peterson et~al.
\newblock High energy gamma-ray imaging using cherenkov cone detection - a
  monte carlo study with application to a compton camera system.
\newblock In {\em 2012 IEEE Nuclear Science Symposium and Medical Imaging
  Conference Record (NSS/MIC)}, pages 3246--3253, Oct 2012.

\bibitem{ref:Roellinghoff}
F.~Roellinghoff et~al.
\newblock Design of a compton camera for 3d prompt-γ imaging during ion beam
  therapy.
\newblock {\em Nuclear Instruments and Methods in Physics Research Section A:
  Accelerators, Spectrometers, Detectors and Associated Equipment}, 648:S20 --
  S23, 2011.

\bibitem{ref:ota}
R.~Ota et~al.
\newblock Cherenkov radiation-based three-dimensional position-sensitive pet
  detector: A monte carlo study.
\newblock {\em Medical Physics}, 45(5):1999--2008, 2018.

\bibitem{ref:peterson2016}
A.~H. Walenta et~al.
\newblock Gamsim -- a windows-based simulation tool for gamma-ray detector
  development.
\newblock In {\em 2016 IEEE Nuclear Science Symposium, Medical Imaging
  Conference and Room-Temperature Semiconductor Detector Workshop
  (NSS/MIC/RTSD)}, pages 1--8, Oct 2016.

\bibitem{ref:Cherenkov34}
P.~A. Cherenkov.
\newblock {Visible emission of clean liquids by action of γ radiation}.
\newblock {\em Doklady Akademii Nauk SSSR}, 2:451+, 1934.

\bibitem{ref:kolanoski}
{H. Kolanoski and N. Wermes}.
\newblock {\em Particle Detectors}.
\newblock Springer Spektrum, 2016.

\bibitem{ref:sultanova}
{N. Sultanova} et~al.
\newblock Dispersion properties of optical polymers.
\newblock {\em Acta Physica Polonia A}, 116(4):585, 2009.

\bibitem{ref:hamamatsu}
{Hamamatsu Photonics K.K.}
\newblock {\em MPPC (Multi-Pixel Photon Counter), S13360 series}, 2016.

\bibitem{ref:estar}
M.~J. Berger et~al.
\newblock Nist standard reference database 124.
\newblock National Institute of Standards and Technology, 2017.
\newblock {Gaithersburg, MD}.

\bibitem{ref:tpx}
{Mitsui Chemicals, Inc.}
\newblock {\em {TPX Transparent Polymer X}}, 2018.

\bibitem{ref:stic2014}
T.~Harion et~al.
\newblock Stic — a mixed mode silicon photomultiplier readout asic for
  time-of-flight applications.
\newblock {\em Journal of Instrumentation}, 9(02):C02003, 2014.

\end{thebibliography}





\end{document}